\documentclass[journal=jpclcd,manuscript=letter,layout=twocolumn]{achemso}
\usepackage{amsfonts,amsmath,amssymb,mathrsfs}
\usepackage[english]{babel}
\usepackage{booktabs}
\usepackage{cancel}
\usepackage[font=small]{caption}
\usepackage{dsfont}
\usepackage{etoolbox} 
\usepackage{float}
\usepackage[T1]{fontenc}
\usepackage{graphicx}
\usepackage[utf8]{inputenc}
\usepackage[hidelinks]{hyperref}
\usepackage{physics}
\usepackage{mathtools}
\usepackage{setspace}
\usepackage[absolute,overlay]{textpos}
\usepackage{tcolorbox}
\usepackage{svg}
\usepackage[normalem]{ulem}
\usepackage{widetext}
\usepackage{xcolor}
\usepackage{xr}
\usepackage{xspace}
\usepackage{kbordermatrix}
\renewcommand{\kbldelim}{(}
\renewcommand{\kbrdelim}{)}

\graphicspath{{Figures/}}

\usepackage[version=3]{mhchem} 

\makeatletter
\SectionNumbersOn
\makeatother

\newcommand{\deriv}[2]{\frac{\partial #1 }{\partial #2}}
\newcommand{\derivv}[2]{\frac{\partial^2 #1 }{\partial #2 ^2}}
\newcommand{\derivw}[3]{\frac{\partial^2 #1}{\partial #2 \partial #3}}

\newcommand*{\isDefinedAs}{\coloneqq}

\newcommand*{\C}{\Bar{C}}
\newcommand*{\U}{\Bar{U}}
\newcommand*{\G}{\Bar{G}}
\newcommand*{\E}{\Tilde{E}}

\newcommand*{\Enx}{\bar{E}}
\newcommand*{\deltaPlus}{\delta^+}
\newcommand*{\deltaMinus}{\delta^-}

\newcommand*{\xv}{\mathbf{x}}
\newcommand*{\rv}{\mathbf{r}}
\newcommand*{\ext}{\text{ext}}
\newcommand*{\Nbas}{N}
\newcommand*{\paper}{letter\xspace}

\newcommand*{\Integers}{\mathbb{Z}}

\newcommand*{\e}{\textrm{e}}
\newcommand*{\im}{\textrm{i}}
\newcommand*{\ud}{\mathrm{d}}

\DeclareMathOperator{\sinc}{sinc}

\newcommand*{\binteg}[3]{\int^{\mathrlap{#3}}_{\mathrlap{#2}}\ud{#1}\,}

\let\oldmaketitle\maketitle
\let\maketitle\relax

\author{Nicolas G. Cartier}
\author{Klaas J. H. Giesbertz}
\email{k.j.h.giesbertz@vu.nl}
\affiliation[VU-Amsterdam]{Department of Chemistry \& Pharmaceutical Sciences and Amsterdam Institute of Molecular and Life Sciences (AIMMS), Faculty of Science, Vrije Universiteit, 1081HV Amsterdam, The Netherlands}
\title{Impact of Parametrizations of the One-Body Reduced Density Matrix on the Energy Landscape}
\abbreviations{}

\begin{document}


\begin{tocentry}
    \vspace*{\fill}
    \begin{tcolorbox}[colframe=white,colback=white,arc=0mm,enlarge left by=-0.2cm,enlarge right by=0.2cm,width=5.5cm,left=0cm,right=0cm,top=0.1cm,bottom=0.1cm,enlarge top by=-0.2cm]
    \begin{center}
        $\gamma$-space \\
        \includesvg[width=0.45\textwidth]{Figures/Muller_H2_sto3g_gamma_space_sf.svg}\\
        \vspace{-0.3cm}
        \includesvg[width=0.45\textwidth]{Figures/Muller_H2_sto3g_x_space_exp_sf.svg}
        \includesvg[width=0.45\textwidth]{Figures/Muller_H2_sto3g_x_space_Cayley_sf.svg}\\
        $U=\exp(X)$\hspace{0.3cm}$U=(I-X)(I+X)^{-1}$
    \end{center}
    \end{tcolorbox}
    \vspace*{\fill}
\end{tocentry}
\twocolumn[
    \begin{@twocolumnfalse}
    \oldmaketitle
    \begin{abstract}
    Many electronic structure methods rely on the minimization of the energy of the system with respect to the one-body reduced density matrix (1-RDM). To formulate a minimization algorithm, the 1-RDM is often expressed in terms of its eigenvectors via an orthonormal transformation and its eigenvalues.~This transformation drastically alters the energy landscape. Especially in 1-RDM functional theory this means that the convexity of the energy functional is lost. We show that degeneracies in the occupation numbers can lead to additional critical points which are classified as saddle points. Using a Cayley or Householder parametrization for the orthonormal transformation, no extra critical points arise. In case of Given's rotations or the exponential, additional critical points can arise, which are of no concern in practical minimization. These findings provide an explanation for the success of recent minimization procedures using second-order information.
    \end{abstract}
    \end{@twocolumnfalse}
]



Most cost effective electronic structure calculations involve the optimization of some energy functional w.r.t.\ the one-body reduced density matrix (1-RDM). Most notable are Hartree--Fock (HF),\cite{Hartree1928, Dirac1926, Fock1930} Kohn--Sham (KS) density functional theory (DFT) \cite{HohenbergKohn1964, KohnSham1965, DreizlerGross2012,Sarmah2017,KohnBecke1996,BearendsGritsenko1997,MardirossianHeadGordon2017,OrioPantazis2009,Bartolotti1981,HaunschildBarth2016,TealeHelgaker2022} and 1-RDM functional theory (RDMFT).\cite{Gilbert1975, Levy1979, Valone1980} During optimization, one needs to ensure that physical 1-RDMs are generated, i.e.\ they should be hermitian and the occupation numbers $n_i$ (eigenvalues of the 1-RDM) should conserve the number of electrons $\sum_i n_i=N_e$ and satisfy the Pauli constraints $0 \leq n_i \leq 1$.\cite{Lowdin1955, Coleman1963} In Hartree--Fock theory, the 1-RDM should even be idempotent $n_i \in \{0,1\}$, a restriction which is also often imposed on the Kohn--Sham system in DFT.
In RDMFT, the constraints on the occupation numbers naturally lead to the natural optimization strategy to use an explicit parametrization of the 1-RDM in its spectral decomposition (diagonal form), since that allows one to easily enforce the Pauli constraints on the occupation numbers. Moreover, many approximate functionals in RDMFT are formulated in terms of its spectral decomposition, making the perspective of optimizing w.r.t.\ this spectral decomposition even more attractive.\cite{GoedeckerUmrigar1998,GritsenkoPernal2005, PirisMatxain2010,PirisLopez2011, Piris2017,Piris2021, Piris2023} 

The price one has to pay, however, is that the Valone interaction functional (defined later in ~\eqref{eq:Valone}), which is a convex functional of the 1-RDM,\cite{PhD-Giesbertz2010, ZumbachMaschke1985} is not convex anymore when written in terms of the natural occupation numbers $n_i$ and the natural orbitals (NOs) $\phi_i$ (eigenvalues and corresponding eigenfunctions of the physical 1-RDM), making the numerical optimization of the total energy harder to converge.
Similarly, the HF energy functional is a simply quadratic functional in terms of the 1-RDM, but becomes a more complicated functional in terms of its spectral decomposition. Also in KS-DFT we expect a less regular energy functional.
So, in all these cases, the energy landscape to optimize over will become more irregular than the original one and additional critical points may result from the parametrization, which are especially concerning if they are local minima.

In this work we will show that additional critical points can indeed be created by the spectral parametrization of the 1-RDM if there are degenerate natural occupation numbers, but that the second-order derivative test is able to classify these critical points as saddle points.
Additional critical points are generated, depending on the type of parametrization one uses for the orthonormal variations of the eigenfunctions. However, we argue that these critical points do not cause problems in typical optimization strategies, so we have not investigated these further. In the next paragraphs we will briefly introduce RDMFT and sketch some of the approaches that have been designed to optimize the energy in this framework, which also allows us to introduce our notations.

The main motivation for interest in RDMFT is that DFT fails even qualitatively for strongly correlated systems.\cite{CohenMoris-Sanchez2008,CohenMoriSanchez2012,BallySastry1997,BraidaPhilippe1998,OssowskiBoyer2003,GruningGritsenko2003,RuzsinskyPerdew2006,Burke2012,WagnerBaker2014,PerdewRuzsinsky2021} RDMFT provides a promising alternative, capable of accurately treating strongly correlated systems while maintaining a reasonable scaling.\cite{PirisOtto2003,Mazziotti2007,PernalGiesbertz2016,KamilShade2016,LathiotakisMarques2008,PirisMatxain2010,RodriguezMayorgaRamosCordoba2017,Piris2023} 
The foundation of RDMFT followed roughly the one of DFT in history. First a Hohenberg--Kohn type argument was put forward by Gilbert \cite{Gilbert1975} and subsequently the constrained search formulation over pure states was put forward by Levy.\cite{Levy1979} The pure state version is more problematic in RDMFT than in DFT, since the pure state $N$-representability conditions are quite involved.\cite{AltunbulakKlyachko2008,SchillingPhD2014} Therefore Valone extended the constrained search to mixed states\cite{Valone1980}
\begin{align}\label{eq:Valone}
W[\gamma] \isDefinedAs \min_{\{w_P,\Psi_P\} \to \gamma} \sum_Pw_P\langle\Psi_P \vert \hat{W} \vert \Psi_P\rangle,
\end{align}
where $\hat{W}$ is the electron-electron interaction operator and $\{w_P,\Psi_P\}$ are a set of states $\Psi_P$ and respective weight $w_P$ (satisfying $w_P\geq 0$ and $\sum_P w_P=1$) generating $\gamma$. Let us stress that the Gilbert, Levy and Valone functionals coincide on the pure state $v$-representable domain, i.e.\ the domain of physical interest.~\cite{Schilling2018,GritsenkoPernal2019} The total electronic energy then writes
\begin{align}\label{eq:RDMFT_E}
    E[\gamma] &\isDefinedAs -\frac{1}{2}\int\bigl[\Delta_{\rv}\gamma(\xv,\xv')\bigr]_{\xv' = \xv}d \xv\notag \\
    &+ \iint \gamma(\xv,\xv')v_{\ext}(\xv',\xv) d\xv d\xv' + W[\gamma],
\end{align}
with $\Delta_{\rv}$ Laplacian w.r.t.\ the position $\rv$, $\xv = \rv\sigma$ a spin-position and $v_{\ext}$ the external (non-local) potential. 

\begin{figure*}[t]
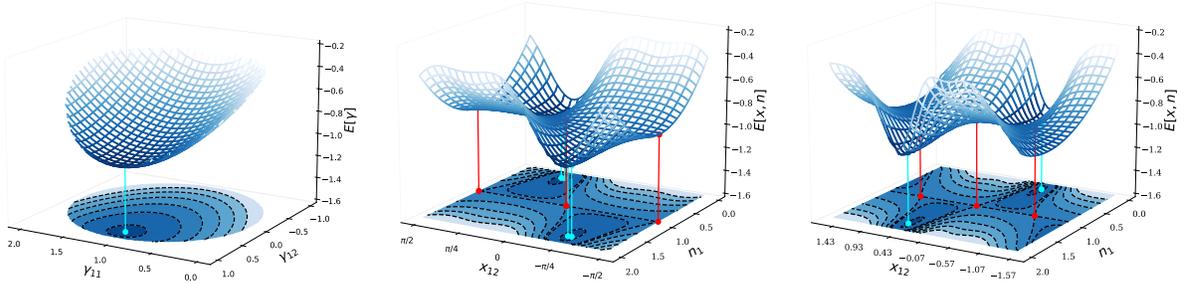

    \begin{center}
        \includesvg[width=0.3\textwidth]{Figures/Muller_H2_sto3g_gamma_space.svg}
        \includesvg[width=0.3\textwidth]{Figures/Muller_H2_sto3g_x_space_exp.svg}
        \includesvg[width=0.3\textwidth]{Figures/Muller_H2_sto3g_x_space_Cayley.svg}
	\caption{Energy of the Müller functional for \ce{H2} in the minimal basis, w.r.t.\ the 1-RDM $\gamma$ (left panel) and the variables prametrizing $\gamma$, $x_{12}$ and $n_1$ as in~\eqref{eq:1rdm_param}, using an exponential (central panel) and the Cayley parametrization (right panel) for the orthonormal matrix. Note that in the present case, of only two orbitals, the Givens and exponential as well as the Cayley and Householder parametrisation are identical up to a sign and so than we show only the two distinct ones. The minima are indicated by cyan dots and the saddle point by a red dot. Note that $E[\gamma]$ is convex while $E[x_{12},n_{1}]$ is not.}	
        \label{fig:E_convx_gamma_vs_NU_space}
    \end{center}
\end{figure*}

Apart from some explorations,\cite{CancesPernal2008,VladajMarecat2024} people do not directly perform the minimization of $E$ w.r.t.\ the 1-RDM but w.r.t.\ some variables $x_i$ parametrizing the natural occupation numbers and natural orbitals. The functional to minimize, $E[\{x_i\}]$ does then not have to be convex, in general. An example were the convexity is lost is presented in Fig.~\ref{fig:E_convx_gamma_vs_NU_space} for the Müller functional\cite{Muller1984,BuijsePhD1991,BuijseBaerends2002}
\begin{align}\label{eq:muller_func}
    W^{\text{MBB}}[\gamma]=& \frac{1}{2}\sum_{ijkl}\Big( \gamma_{ij}[ij|kl]\gamma_{kl}\notag\\
    &-\sqrt{\gamma}_{ij}[il|kj]\sqrt{\gamma}_{kl}\Big),
\end{align}
with 
\begin{equation}
    [ij|kl] \isDefinedAs \iint \frac{\chi_i^*(\xv)\chi_j(\xv)\chi_k^*(\xv')\chi_l(\xv')}{|\rv-\rv'|}d\xv d\xv',
\end{equation}
in the basis $\{\chi\}$.

Many parametrizations have been considered for both natural occupation numbers and NOs.\cite{Pernal2005,CancesPernal2008,BaldsiefenGross2013,PirisMitxelena2021,LemkeKussmann2022,Piris2023} For the natural occupation numbers, one often uses $n_i(x_i)=\cos(x_i)^2$\cite{Gilbert1975,HerbertHarriman2003} with $x_i\in\mathds{R}\forall i$, $n_i(x_i)=\bigl(1+\exp(x_i-\mu)\bigr)^{-1}$\cite{BaldsiefenGross2013} or $n_i(x_i)=\frac{1}{2}\bigl(\erf(x_i+\mu)+1\bigr)$\cite{YaoFang2021,YaoSu2024} ($\mu\in\mathds{R}$) to impose the constraints on the boundaries of $n_i$. The last two can also impose the constraint on the trace of the 1-RDM by varying $\mu$, and do not create additional critical points at a given $\mu$.\cite{YaoFang2021}
Since a parametrization of the form $n_i(x_i)=f(x_i)\forall i$, in the case of $f$ bijective or periodic with a single minimum per period cannot create local minima (see Appendix \ref{sec:unic_min_periodic}) and that one can also impose the constraints using Lagrange multipliers\cite{LathiotakisHelbig2005,LathiotakisHelbig2007,TheophilouLathiotakis2015} or other constraint minimization methods,\cite{CohenBaerends2002} we will consider $n_i$ i.e.\ the occupations themselves as the variables. Regarding the NOs, one usually wants to keep them orthonormal and often employs a unitary parametrization to do so, more specifically an orthonormal parametrization (since one usually works with real orbitals\footnote{restriction taken in this \paper as well}) or iteratively diagonalizes a generalized Fock matrix,\cite{Pernal2005,PirisUgalde2009,LemkeKussmann2022} which is effectively the same in the context of a gradient descent. We will therefore consider the following typical parametrization of the 1-RDM in this \paper
\begin{equation}
    \label{eq:1rdm_param}
    \gamma_{\mu\nu}(n,x) = \sum_{k} U_{\mu k}(x)n_k U_{k\nu}(x)^T, 
\end{equation}
where $U(x)$ is an orthonormal matrix which represents the NOs in some orthonormal basis of $N$ orbitals and depends on a set of variables $x$.

The most popular form for $U$ is the exponential of a skew-symmetric matrix\cite{ShepardBrozell2015,DouadyEllinger1980,CioslowskiPernal2001,Sun2017,LehtolaBlockhuys2020} but others are possible. We will consider additionally the parametrization via a Cayley transform,\cite{Cayley1846,Weyl1946,ShepardBrozell2015} a product of Givens rotations\cite{Givens1958,GeorgeLiu1987,GolubVanLoan2012,ShepardBrozell2015} and of Householder reflections.\cite{Householder1958,GeorgeLiu1987,GolubVanLoan2012,ShepardBrozell2015}

The parametrization then allows us to write the energy as a functional of the parameters $n,x$ via the straightforward composition
\begin{align}
\Enx[n,x] \isDefinedAs E\bigl[\gamma(n,x)\bigr] .
\end{align}
Note that the 1-RDM is symmetric, so in the following we will take $E$ to be only dependent on the upper triangular part of $\gamma$, i.e.\ only the elements $\gamma_{\mu\nu}$ with $\mu \leq \nu$. The variables $x$ for the NO parametrization will be indexed as $x_{ij}$ with $i < j$.


\paragraph{Identification of the critical points}
At a critical point in the parameter space $(n,x)$, we have
\begin{equation}
    \grad_{n,x}\Enx = \bigl(\grad_{n,x}\gamma\bigr)^T\cdot\grad_{\gamma} E =0 ,
\end{equation}
where $\grad_{\gamma}$ is the gradient in the space of the 1-RDM and $\grad_{n,x}$, the gradient in the parameter space ($\grad_{n,x}\gamma$ is therefore the Jacobian of the parametrization~\eqref{eq:1rdm_param}).

The proper critical points in which we are interested are the ones that correspond to $\grad_{\gamma}E=0$, so that they are actual critical points of $E[\gamma]$ (cyan dots in Fig.~\ref{fig:E_convx_gamma_vs_NU_space}). Note that a unique critical point in $E[\gamma]$ can correspond to multiple critical points in $\Enx[n,x]$ due to periodicity in the parametrization, but they will all be equal in the sense that they give the same energy and corresponding 1-RDM.

However, additional critical points can arise, if $\grad_{\gamma}E$ is in the null space of $\grad_{n,x}\gamma$. The first step is thus to determine when $\grad_{n,x}\gamma$ is singular, that is, when we have $\det\left(\grad_{n,x}\gamma\right)=0$.
With the parametrization~\eqref{eq:1rdm_param}, we can show that (see Appendix~\ref{sec:detjac_proof} for details),
\begin{multline}
    \label{eq:detjac}
    \det(\grad_{n,x}\gamma) =
    \det\left(U^T\grad_x U\right)\prod_{p<q}(n_q-n_p).
\end{multline}
We notice that this determinant is 0 when occupation numbers are degenerate
(case of the red dots in Fig.~\ref{fig:E_convx_gamma_vs_NU_space}, recall that $n_2=2-n_1$ for \ce{H2} in minimal basis; an analytical example is also provided in Appendix~\ref{sec:crit_pt_ex}). This is due to the fact that a rotation between the corresponding orbitals is then irrelevant, making the energy independent of $x_{pq}$. So each $k$-fold degeneracy in the natural occupation numbers adds $k(k-1)/2$ dimensions to the null space of the Jacobian. This part of the null space is explicitly constructed in Appendix~\ref{sec:detjac_proof}. In practice we will have numerically a very large null space, since many occupation numbers tend to be very close to each other, especially in a large basis set. So, when close to convergence ($\grad_{n,x}E \approx 0$) $\grad_{\gamma}E$ tends to have a significant component in the (numerical) null space of $\grad_{n,x}\gamma$.

To see whether the orbital parametrization incurs additional critical points,  we have to specify $U$. For a Cayley transform
\begin{equation}\label{eq:cayley_def}
    U(X) = (I-X)(I+X)^{-1}
\end{equation}
with $X_{ij} =x_{ij}\forall i<j, \, X_{ij} =-x_{ij} \forall i>j$ and $X_{ii}=0$. Then, using $\deriv{X^{-1}}{x_{ij}}=-X^{-1}\deriv{X}{x_{ij}}X^{-1}$ and $U=2(I+X)^{-1}-I$, 
\begin{align}
    \deriv{U}{x_{ij}} &= -2(I+X)^{-1}\deltaMinus_{ij} (I+X)^{-1},
\end{align}
where $\bigl(\deltaMinus_{ij}\bigr)_{kl}=\delta_{ik}\delta_{jl}-\delta_{il}\delta_{jk}$. So each element can be written as a $2 \times 2$ determinant
\begin{equation}
    \left(U^T\deriv{U}{x_{ij}}\right)_{\mathrlap{kl}} = -2 \det((I+X)^{-1}_{\{ij\}\{kl\}}) ,
\end{equation}
where $X_{JL}$ refers to the submatrix of $X$ obtained by keeping the rows in $J$ and columns in $L$. Defining $(I+X)^{-(2)}_{ij,kl}\isDefinedAs \det\big( (I+X)^{-1}_{\{ij\}\{kl\}})$,
\begin{align}
    \det(U^T\grad_x U) &= (-2)^{N(N-1)/2}\det((I+X)^{-(2)}) \notag \\
    &= (-2)^{N(N-1)/2}\det(I+X)^{1-N}\notag\\
    &\neq 0 ,
\end{align}
by the Sylvester--Franke theorem.\cite{Price1947} That is, the Cayley parametrization has the advantage not to introduce any additional critical points.

\begin{figure}[t]
    \begin{center}
        \includesvg[width=0.7\textwidth]{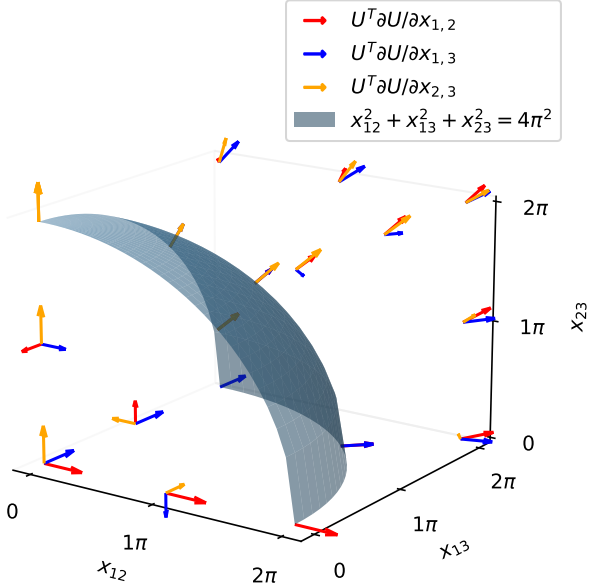}
	\caption{Tangent vectors of the exponential parametrisation $U=\exp(X)$ (with $X^T=-X$) for $N=3$. The vectors align on the sphere $x_{12}^2+x_{13}^2+x_{23}^2=4\pi^2$ since the eigenvalues of $X$ are $\pm\sqrt{-x_{12}^2-x_{13}^2-x_{23}^2}$ and 0.}	
        \label{fig:UdU_exp_vec_field}
    \end{center}
\end{figure}

Similarly, (see Appendix~\ref{sec:detjac_exp}) we can show that, when we take $U=\exp(X)$ ($X$ defined as previously),
\begin{equation}
\det\left(U^T\grad_x U\right)
= \prod_{k<l}\sinc\biggl(\frac{\lambda_k + \lambda_l}{2}\biggr) ,
\end{equation}
where $\lambda_k $ is the imaginary part of the $k^{th}$ eigenvalue of $X$. Thus, an exponential parametrization can have additional critical points for each pair \(\lambda_k + \lambda_l = 2\pi m\) with \(m \in \Integers \setminus \{0\}\). The factor $2\pi$ is related to the fact that the exponential is a many-to-one map modulo $2\pi$ in the eigenvalues. At these singularities some of the tangent vectors to the parametrization of $\gamma$ align and therefore do not span the full $N(N-1)/2$ vector space anymore (see Fig.\ref{fig:UdU_exp_vec_field}).

For an orthonormal parametrization $U=\prod_{k<l}^{\downarrow}G(k,l,x_{kl})$, where $G_{pq}(k,l,x_{kl}) \isDefinedAs \delta_{pq} + \delta_{pq}(\delta_{pk} + \delta_{pl})\bigl(\cos(x_{kl}) - 1\bigr) + (\delta_{pl}\delta_{qk}-\delta_{pk}\delta_{ql})\sin(x_{kl})$ is the Givens rotation of angle $x_{kl}$ in the plane spanned by $\{e_k,e_l\}$ (with $(e_k)_p= \delta_{kp}$) and $\Pi^{\downarrow}_{k<l}$ indicates a product in descending index order, we obtain (see Appendix \ref{sec:detjac_givens})
\begin{multline}
    \det\left(U^T\grad_x U\right) \\
    = (-1)^{N(N-1)/2}
    \prod_{l\geq3}^N\prod_{k=1}^{l-2}\cos(x_{kl})^{l-k-1}.
\end{multline}
Thus, a parametrization with Givens rotations induces critical points for $x_{kl}=\frac{\pi}{2}[\pi]\forall k< l-1$. These angles are the generalization of the gimbal lock in SO(3) to SO($N$), that is, they make the variables $x_{ki}$ and $x_{il}$ degenerate for each $i$.

Finally, using Householder reflections, $U=\prod_{l}^{\downarrow}H(l)$, with $H(l)\isDefinedAs 1-2\frac{v_lv_l^T}{\norm{v_l}^2}$ (with reflections ordered in descending index order),
\begin{equation}
    \label{eq:householder_vect_def}
    (v_l)_k \isDefinedAs
    \begin{cases}
        x_{kl} & k<l \\
        1 & k=l \\
        0 & k>l.
    \end{cases}
\end{equation}
We can then show (see Appendix~\ref{sec:detjac_householder}),
\begin{equation}
    \det\left(U^T\grad_x U\right)= 2^{N(N-1)/2}\prod_{l=2}^N \norm{v_l}^{2(1-l)} \neq0.
\end{equation}
The two problematic parametrizations are thus the exponential and the product of Givens rotations. 

In practice, however, this does not need to be a real issue, since one usually does not recompute $U(x)$ at each iteration, but updates an orthonormal matrix $C$ as 
\begin{equation}
    C^{(n+1)} = C^{(n)}U(x^{(n)})
\end{equation}
at the $(n+1)^{\text{th}}$ iteration, with $C^{(0)}=1$ (in an orthonormal basis). The parametrization~\eqref{eq:1rdm_param} is then rather
\begin{multline}
    \label{eq:1rdm_update_param}        
    \gamma_{\mu\nu}\bigl(n,x^{(n)},C^{(n)}\bigr) \\
    = \sum_{klm} C^{(n)}_{\mu k}U_{kl}(x^{(n)})n_l U_{lm}(x^{(n)})^TC^{(n)T}_{m\nu}.
\end{multline}
In fact, in this scheme we expect $x^{(n)}$ to be close to $0$, since the step length will typically be kept sufficiently under control such that the operator norm (i.e.\ the largest absolute eigenvalue) \(\norm*\big{X^{(n)}}_2 < \pi\) for the exponential or the max norm (the largest absolute matrix element) \(\norm*\big{X^{(n)}}_{\max} < \pi/2\) for the Givens rotations.
The critical points induced by the orthonormal parametrization should therefore play a minor role in practice\footnote{Since, the gradient is evaluated at $x=0$ at each step, the critical points due to the orbital parametrization do not prevent convergence, even if they would correspond to local minima.} and we will focus on the degeneracy of the occupations in the next section.

\paragraph{Nature of the critical points at degenerate natural occupation numbers}
In the previous paragraph, we noticed that when occupations are degenerate, the parametrization can produce a critical point while having $\grad_{\gamma}E\neq0$. We will now show that in this case the critical point is not a minimum.

To do so, we consider the parametrization~\eqref{eq:1rdm_update_param}
and consider the orbital components of partial derivatives of the energy at $x=0$\cite{CartierGiesbertz2024,YaoSu2024}
\begin{subequations}
\begin{align}
    \label{eq:dEdni}
    \deriv{\Enx}{n_i}\biggr\rvert_{\mathrlap{x=0}}
    &\propto \sum_{\mu\leq\nu}\deriv{E}{\gamma_{\mu\nu}}C_{\mu i}C_{\nu i} 
    \propto \frac{1}{2}\E_{ii} , \\
    \label{eq:dEdxij}
    \deriv{\Enx}{x_{ij}}\biggr\rvert_{\mathrlap{x=0}}
    &\propto (n_j-n_i)\sum_{\mu\leq\nu}\deriv{E}{\gamma_{\mu\nu}}
    \bigl( C_{\mu j}C_{\nu i} + C_{\mu i}C_{\nu j}\bigr) \notag \\
    &\propto (n_j-n_i)\E_{ij} ,
\end{align}
\end{subequations}
where we defined  (up to an irrelevant multiplicative constant) 
\begin{align}\label{eq:EtildeDef}
\E_{ij} \isDefinedAs \sum_{\mu\leq\nu}\deriv{E}{\gamma_{\mu\nu}}\big(C_{\mu j}C_{\nu i} + C_{\mu i}C_{\nu j}\bigr).
\end{align}
So \(\grad_{n,x}\Enx = 0\) implies $\E_{ii} = 0$ and $\E_{ij} = 0$ for all non-degenerate $i,j$ pairs, i.e.\ with $n_i \neq n_j$.
If the critical point is also a minimum, we can actually argue that $\E_{ij} = 0$ is also zero. For this, we need the following elements of the hessian
\begin{subequations}
\begin{align}
    \derivw{\Enx}{x_{ij}}{n_{k}}\biggr\rvert_{\mathrlap{x=0}}
    \propto{}& \frac{n_j-n_i}{2}\E_{kkij} + (\delta_{jk}-\delta_{ik})\E_{ij}, \\
    \label{eq:dEdxdx}
    \derivw{\Enx}{x_{ij}}{x_{kl}}\biggr\rvert_{\mathrlap{x=0}} 
		\propto{}&(n_j-n_i)(n_l-n_k)\E_{ijkl} \notag\\*
            & {}- \delta_{ik}(n_i-n_j/2-n_l/2)\E_{jl}\notag\\*
            & {}- \delta_{jl}(n_j-n_i/2-n_k/2)\E_{ik}\notag\\*
		& {}+ \delta_{il}(n_i-n_j/2-n_k/2)\E_{jk}\notag\\*
            & {}+ \delta_{jk}(n_j-n_i/2-n_l/2)\E_{il} ,
\end{align}
\end{subequations}
where we have defined (up to a constant)  
\begin{multline}
\E_{ijkl}\isDefinedAs\sum_{\mathclap{\mu\leq\nu , \kappa\leq\lambda}}\big(C^T_{j\mu}C^T_{i\nu}+C^T_{i\mu}C^T_{j\nu}\big)\derivw{E}{\gamma_{\mu\nu}}{\gamma_{\kappa\lambda}} \notag \\
\cdot \bigl(C_{\kappa l}C_{\lambda k}+C_{\kappa k}C_{\lambda l}\bigr) .
\end{multline}
Now consider the principal minor of $\grad^2_{n,x}\Enx$ involving the derivatives w.r.t.\ $x_{ij}$ and $n_i$ only for $n_i = n_j$
\begin{equation}\label{eq:principalMinor}
\begin{pmatrix}
\derivv{\Enx}{x_{ij}}           &\derivw{\Enx}{x_{ij}}{n_{i}} \\
\derivw{\Enx}{x_{ij}}{n_{k}}    &\derivv{\Enx}{n_i}
\end{pmatrix}
= \begin{pmatrix}
0       &\E_{ij} \\
\E_{ij} &\derivv{\Enx}{n_i}
\end{pmatrix} .
\end{equation}
By Sylvester's criterion, $\grad^2_{n,x}\Enx$ can only be positive semi-definite (which corresponds to a minimum), if the determinant of all principal minors is non-negative.\cite{GhorpadeLimaye2007} Since the determinant of the principal minor in~\eqref{eq:principalMinor} is $-\E_{ij}^2 \leq 0$, the hessian can be positive semi-definite if and only if $\E_{ij} = 0$.

Now we rewrite $\E_{ij}$ as
\begin{align}
\E_{ij} = \sum_{\mu\leq\nu}\deriv{E}{\gamma_{\mu\nu}}\sqrt{1 + \delta_{\mu\nu}}\C_{\mu\nu,ij}\sqrt{1 + \delta_{ij}} ,
\end{align}
where
\begin{equation}
    \C_{\mu\nu,kl}\isDefinedAs\frac{C_{\mu k}C_{\nu l}+C_{\mu l}C_{\nu k}}{\sqrt{(1+\delta_{\mu\nu})(1+\delta_{kl})}}.
\end{equation}
As we show in appendix~\ref{sec:detjac_proof}, $\C$ ($\U$ in the appendix) is unitary, so from $\E = 0$ we can conclude that $\grad_{\gamma}E = 0$ as well.

The degeneracy of occupations can therefore only induce a saddle point.\footnote{replacing `positive' by `negative' in the above demonstration we show that we cannot have a maximum either} Since these critical points necessarily correspond to an indefinite hessian in the parameter space, second-order algorithms are sufficient to escape from these saddle points.

To conclude, we have shown that using the spectral decomposition of the 1-RDM as parametrization can introduce additional critical points for degenerate natural occupation numbers. Additionally, we have investigated whether additional critical points are induced by the parametrization of the NOs via orthonormal matrices for four different parametrizations: in terms of an exponential, Cayley transform, Givens rotations and Householder reflections. In particular, we have shown that the Householder reflections and the Cayley transform do not introduce additional critical points, but the exponential and Given's rotation can do this for special parameter values. However,
none of these parametrizations induces additional critical points around the reference point ($x = 0$), which is the most relevant point for practical calculations, since at $x=0$ the evaluation of the derivatives of the orthonormal parametrizations becomes particularly simple.

So the only additional critical points of real concern are if occupations are degenerate. However, we have proven that this case corresponds to an indefinite hessian in the parameter space, so a saddle point. 
Many occupation numbers are numerically degenerate, leading to a very large numerical null space of the Jacobian. So, a significant part of the gradient in the 1RDM space barely contributes to the gradient in parameter space. This provides an explanation why gradient based minimization procedures on the spectral parametrization of the 1RDM exhibit very slow convergence and cannot give tight convergence, especially in large basis sets.\cite{Piris2023,PirisMitxelena2021}
The fact that these additional stationary points correspond to saddle points and not local minima provides an explanation for the recent success of second-order optimization methods.\cite{CartierGiesbertz2024,YaoSu2024}

Finally, although our work shows that optimizing w.r.t.\ occupations and NO variables does not incur local minima, the possible presence of multiple critical points suggests that work in the direction of a minimization w.r.t.\ $\gamma$ is a sensible alternative to improve convergence, as suggested by preliminary results.\cite{CancesPernal2008,VladajMarecat2024} 

In particular, the development of an accurate convex functional of $\gamma$ would allow for quadratic convergence, guaranteed for convex problems with a suitable algorithm, which is not generally achievable when optimizing w.r.t.\ $x$ and $n$.  

In the future, we wish to also consider the complex case. Preliminary studies in our group indicate that additional critical points arise for complex orbitals, which requires attention.

Since in multi-configuration self-consistent field (MCSCF) orbital optimization is also used, it would be interesting to apply these ideas to MCSCF type wave functions.
A significant difference in the MCSCF setting is that we do not have natural occupation numbers to optimize, but instead the (non-redundant) expansion coefficients of the wave function.
A first attempt should probably be made with the most well-known and simplest variant complete active space self-consistent field (CASSCF).\cite{SiegbahnHeiberg1980,JensenJorgensen1984,Roos1987,SunZhang2020}


\begin{acknowledgement}
The authors thank the The Netherlands Organization for Scientific Research, NWO, for its financial support under Grant No.~OCENW.KLEIN.434 and Vici Grant No.~724.017.001.
\end{acknowledgement}


\appendix

\section{Proof of the Expression for the Determinant of the Jacobian}
\label{sec:detjac_proof}

We want to derive equation~\eqref{eq:detjac}, expression of the determinant of the Jacobian for the parametrization~\eqref{eq:1rdm_param}.
To do so, it is convenient to first define
\begin{equation}
    \label{eq:Ubar_def}
    \U_{\mu\nu,kl}\isDefinedAs\frac{U_{\mu k}U_{\nu l}+U_{\mu l}U_{\nu k}}{\sqrt{(1+\delta_{\mu\nu})(1+\delta_{kl})}}.
\end{equation}
This `super' matrix is unitary under summation over unique index-pairs, e.g.\ for $k \leq l$
\begin{widetext}
    \begin{align}
    \sum_{k\leq l}\U_{\mu\nu,kl}\U^T_{kl,\kappa\lambda}
    &= \sum_{k\leq l}(1+\delta_{kl})^{-1}(1+\delta_{\mu\nu})^{-1/2}(1+\delta_{\kappa\lambda})^{-1/2}(U_{\mu k}U_{\nu l}+U_{\mu l}U_{\nu k})(U^T_{k\kappa}U^T_{l\lambda}+U^T_{k\lambda}U^T_{l\kappa}) \notag \\
    &=(1+\delta_{\mu\nu})^{-1/2}(1+\delta_{\kappa\lambda})^{-1/2}\sum_{kl}U_{\mu k}U_{\nu l}(U^T_{k\kappa}U^T_{l\lambda}+U^T_{k\lambda}U^T_{l\kappa}) \notag \\
    &=(1+\delta_{\mu\nu})^{-1/2}(1+\delta_{\kappa\lambda})^{-1/2}(\delta_{\mu\kappa}\delta_{\nu\lambda}+\delta_{\mu\lambda}\delta_{\nu\kappa}) ,
    \end{align}%
\end{widetext}%
so $\U\U^T = 1$ by taking in account $\mu\leq\nu, \kappa\leq\lambda$. Similarly $\U^T\U=1$, and thus $\U$ is an orthonormal matrix.

We also define 
\begin{equation}
    \label{eq:bardelta_def}
    \deltaPlus_{ij,kl}=\delta_{ik}\delta_{jl}+\delta_{il}\delta_{jk},
\end{equation}
diagonal matrix with 2 on the diagonal for $k=l$ (and 1 otherwise), so $\det(\deltaPlus)=2^\Nbas$.

This allows us to rewrite~\eqref{eq:1rdm_param} as
\begin{align}
\gamma_{\mu\nu} 
&= \sum_{k \leq l}(\deltaPlus_{\mu\nu,\mu\nu})^{1/2}\U_{\mu\nu,kl} \; (\deltaPlus_{kl,kl})^{-1/2} \; n_k\delta_{kl} \notag \\
&= \bigl((\deltaPlus)^{1/2}\U(\deltaPlus)^{-1/2}n\bigr)_{\mu\nu} ,
\end{align}
where in the last step $n_k\delta_{kl}$ is considered as a vector.
Working out the partial derivatives of the 1-RDM w.r.t.\ the natural occupation numbers we readily find
\begin{align}
\deriv{\gamma_{\mu\nu}}{n_k}
&= \bigl((\deltaPlus)^{1/2}\U(\deltaPlus)^{-1/2}\bigr)_{\mu\nu,kk} .
\end{align}
So we get a particular simple expression, if we multiply the Jacobian from the left by the non-singular \((\deltaPlus)^{1/2}\U^T(\deltaPlus)^{-1/2}\) and obtain
\begin{equation}
    \left((\deltaPlus)^{1/2}\U^T(\deltaPlus)^{-1/2}\deriv{\gamma}{n_k} \right)_{pq} = \delta_{pk}\delta_{qk}.  
\end{equation}
The determinants are now related as
\begin{equation}
\det((\deltaPlus)^{-1/2}\U^T(\deltaPlus)^{1/2}\grad_{n,x}\gamma)
= \det(\grad_{n,x}\gamma) .
\end{equation}
Also for the derivatives w.r.t.\ the NO parameters, this premultiplication simplifies the expression considerably

\begin{widetext}
    \begin{align}
    \left((\deltaPlus)^{1/2}\U^T(\deltaPlus)^{-1/2}\deriv{\gamma}{x_{ij}} \right)_{\!\!\mathrlap{pq}} &= 
    \sum_{\mu\leq\nu,k} (1+\delta_{\mu\nu})^{-1}(U^T_{p\mu}U^T_{q\nu}+U^T_{p\nu}U^T_{q\mu})\biggl(\deriv{U_{\mu k}}{x_{ij}}U_{\nu k}+U_{\mu k}\deriv{U_{\nu k}}{x_{ij}}\biggr)n_k \notag \\
    &=\sum_{\mu\nu,k}U^T_{p\mu}U^T_{q\nu}\biggl(\deriv{U_{\mu k}}{x_{ij}}U_{\nu k}+U_{\mu k}\deriv{U_{\nu k}}{x_{ij}}\biggr)n_k \notag \\
    &=(n_q-n_p)\left(U^T\deriv{U}{x_{ij}}\right)_{pq} ,
    \end{align}%
\end{widetext}%
where we used that $\deriv{U^T}{x_{ij}}U + U^T\deriv{U}{x_{ij}}=0$. Thus, denoting $\Delta n_{pq}\isDefinedAs n_q-n_p$ 
\begin{multline}
    \det(\grad_{n,x}\gamma) =
    \begin{vmatrix}
        I & 0\\
        0 & \Delta n\, U^T\grad_x U
    \end{vmatrix} \\
    = \det\left(U^T\deriv{U}{x}\right)\prod_{p<q}(n_q-n_p).
\end{multline}
We can then compute $\det\left(U^T\deriv{U}{x}\right)$ for a given orthonormal parametrization.

In the case of an occupation degeneracy $n_p=n_q$, the vectors in the null space of \((\deltaPlus)^{1/2}\U^T(\deltaPlus)^{-1/2}\grad_{n,x}\gamma\) are the vectors of the form \(a\deltaPlus_{pq}\) with $a\in\mathds{R}$.
Thus the null space of $\grad_{n,x}\gamma$, \(\ker_{\text{degen}}\{\grad_{n,x}\gamma\}\), caused by a set of degeneracies is given by their span
\begin{multline}
\ker_{\text{degen}}\{\grad_{n,x}\gamma\} \\
=
\text{span}\bigl(\bigl\{ (\deltaPlus)^{1/2}\U(\deltaPlus)^{-1/2}\deltaPlus_{pq}:
n_p = n_q\bigr\}\bigr) .
\end{multline}

\section{Jacobian of the Exponential}\label{sec:detjac_exp}
We consider the case $U(x) = \exp(X)$ with $X^T=-X$ and, using a well known expression for the derivative of a exponentiated matrix\cite{Brewer1977}
\begin{align}
    U^T\deriv{U}{x_{ij}}&= \binteg{\alpha}{0}{1} \bigl[\exp\bigl(-(1-\alpha)X\bigr)\notag\\
    &\cdot \deltaMinus_{ij}\exp\bigl((1-\alpha)X\bigr)\bigr].
\end{align}
We consider the eigenvalue decomposition of $X$, $U_{pq}=\sum_kV_{pk}e^{\im\lambda_k}V_{kq}^{\dagger}$, with $\lambda_k\in\mathds{R}$, imaginary part of the $k^{th}$ eigenvalue of $X$ and $V_p$, $p^{th}$ eigenvector.
\begin{equation}
    \label{eq:init_jac_exp}
    \left(U^T\deriv{U}{x_{ij}}\right)_{\mathrlap{pq}} = \sum_{kl}\Upsilon_{kl}V_{pk}V^{\dagger}_{lq}(V^{\dagger}\deltaMinus_{ij}V)_{kl} ,
\end{equation}
where changing the integration variable to $\beta=1-\alpha$
\begin{align}
    \Upsilon_{kl} \isDefinedAs{}& \binteg{\beta}{0}{1} \e^{\im\beta}(\lambda_l-\lambda_k) \notag \\
          {}={}&\begin{cases}
              \frac{\e^{\im(\lambda_l-\lambda_k)}-1}{\im(\lambda_l-\lambda_k)} & \lambda_l\neq\lambda_k \\
              1 & \lambda_l=\lambda_k
          \end{cases} .
\end{align}

Since $X$ is skew-symmetric $\forall k\; \exists \Bar{k} \text{ s.t.\ } \lambda_k=-\lambda_{\Bar{k}}$ and $V_{l\Bar{k}}=V_{lk}^* = V_{kl}^{\dagger}$. Then~\eqref{eq:init_jac_exp} can be rewritten as
\begin{align}
    \left(U^T\deriv{U}{x_{ij}}\right)_{\mathrlap{pq}}
    &= \sum_{kl}\Upsilon_{k\Bar{l}}V_{pk}V^{\dagger}_{\Bar{l}q}(V^{\dagger}\deltaMinus_{ij}V)_{k\Bar{l}} \notag \\
    &= \sum_{k < l}\Bigl(\Upsilon_{k\Bar{l}}V_{pk}V^{\dagger}_{\Bar{l}q}(V^{\dagger}\deltaMinus_{ij}V)_{k\Bar{l}} \notag \\
    &\hphantom{{}=\sum_{k < l}\Bigl(}
    + \Upsilon_{l\Bar{k}}V_{pl}V^{\dagger}_{\Bar{k}q}(V^{\dagger}\deltaMinus_{ij}V)_{l\Bar{k}}\Bigr), \notag \\
\end{align}
using that $(V^{\dagger}\deltaMinus_{ij}V)_{kl}=2\Re[V^{\dagger}_{ki}V_{jk}]$ so that $\exists m\in\mathds{R}$, $M\in\mathds{R}$ s.t. $\delta_{pq}m\leq\sum_k\Upsilon_{kk} V_{pk}V^{\dagger}_{kq}(V^{\dagger}\deltaMinus_{ij}V)_{kk} \leq\delta_{pq}M=0$ for $p<q$. Introducing 
\begin{align}
V^{(2)}_{pq,kl}
\isDefinedAs V_{pk}V_{ql} - V_{pl}V_{qk}
=(V\deltaMinus_{kl}V^{\dagger})_{p\Bar{q}},
\end{align}
and using $\Upsilon_{k\Bar{l}}=\Upsilon_{l\Bar{k}}$ and $(V^{\dagger}\deltaMinus_{ij}V)_{k\Bar{l}} = -(V^{\dagger}\deltaMinus_{ij}V)_{l\Bar{k}}$ 
\begin{align}
    \left(U^T\deriv{U}{x_{ij}}\right)_{\mathrlap{pq}}
    = \sum_{k<l}V^{(2)}_{pq,kl}\Upsilon_{k\Bar{l}}V^{(2)\dagger}_{kl,ij} .
\end{align}
Defining $\Bar{\Upsilon}_{pq,kl} \isDefinedAs \delta_{pk}\delta_{ql}\Upsilon_{k\Bar{l}}$,
\begin{equation}
    \det\left(U^T\grad_x U\right) = \det(V^{(2)}\Bar{\Upsilon} V^{(2)\dagger}).
\end{equation}
By the Sylvester--Franke theorem $\det(V^{(2)})=\det(V)^{N-1}=1$ and hence
\begin{align}
    \label{eq:upsilon_prod}
    \det\left(U^T\grad_x U\right)&= \prod_{k<l}\Upsilon_{k\Bar{l}}.
\end{align}
To work out the remaining product, we take the eigenvalues in increasing order and define the sets \(\mathcal{L}^\pm = \{\lambda_k : \pm\lambda_k > 0\}\) and \(\mathcal{L}^0 = \{\lambda_k : \lambda_k = 0\}\). The contribution when $k,l$ are both in $\mathcal{L}^\pm$ or $\mathcal{L}^0$ is

\begin{align}
\prod_{\mathclap{k < l \in \mathcal{L}^-}}&\Upsilon_{k\bar{l}}\cdot
\prod_{\mathclap{k < l \in \mathcal{L}^0}}\Upsilon_{k\bar{l}}\cdot
\prod_{\mathclap{k < l \in \mathcal{L}^+}}\Upsilon_{k\bar{l}}
= \prod_{\mathclap{k < l \in \mathcal{L}^+}}\Upsilon_{k\Bar{l}}\Upsilon_{\Bar{k}l} \notag \\
&= \prod_{\mathclap{k < l \in \mathcal{L}^+}}\frac{\e^{-\im(\lambda_l + \lambda_k)} - 1}{-\im(\lambda_l + \lambda_k)} \cdot 
\frac{\e^{\im(\lambda_l + \lambda_k)} - 1}{\im(\lambda_l + \lambda_k)} \notag \\
&= \prod_{\mathclap{k < l \in \mathcal{L}^+}}
\Bigl(\sinc\bigl((\lambda_k+\lambda_l)/2\bigr)\Bigr)^2 ,
\end{align}

where we used that the contribution from the zero eigenvalues is 1 to the product. The contributions from $\lambda_k < 0$ and $\lambda_l > 0$ become
\begin{align}
\prod_{\mathclap{\substack{k \in \mathcal{L}^- \\ l \in \mathcal{L}^+}}}\Upsilon_{k\bar{l}}
&= \prod_{\mathclap{k < l \in \mathcal{L}^+}}
\Upsilon_{\bar{k}\bar{l}}\Upsilon_{\bar{l}\bar{k}} 
= \prod_{\mathclap{k < l \in \mathcal{L}^+}}
\Upsilon_{lk}\Upsilon_{kl}
\notag \\
&= \prod_{\mathclap{k < l \in \mathcal{L}^+}}\frac{\e^{\im(\lambda_l - \lambda_k)} - 1}{\im(\lambda_l - \lambda_k)}\frac{\e^{-\im(\lambda_l - \lambda_k)} - 1}{-\im(\lambda_l - \lambda_k)} \notag \\
&= \prod_{\mathclap{k < l \in \mathcal{L}^+}}\Bigl(\sinc\bigl((\lambda_l - \lambda_k)/2\bigr)\Bigr)^2 .
\end{align}
The cross contributions with the $\mathcal{L}^0$ (if non empty) contribute as
\\

\begin{align}
\prod_{\mathclap{\substack{k \in \mathcal{L}^- \\ l \in \mathcal{L}^0}}}\Upsilon_{k\Bar{l}}\cdot
\prod_{\mathclap{\substack{k \in \mathcal{L}^{+} \\ l \in \mathcal{L}^{0}}}}\Upsilon_{k\Bar{l}}
&= \prod_{\mathclap{\substack{k \in \mathcal{L}^+ \\ l \in \mathcal{L}^0}}}\Upsilon_{\Bar{k}l}\Upsilon_{kl} \notag \\
&= \prod_{\mathclap{\substack{k \in \mathcal{L}^+ \\ l \in \mathcal{L}^0}}}\frac{\e^{\im\lambda_k} - 1}{\im\lambda_k} \cdot 
\frac{\e^{-\im\lambda_k} - 1}{-\im\lambda_k} \notag \\
&= \prod_{\mathclap{k \in \mathcal{L}^+}}
\Bigl(\sinc(\lambda_k/2)\Bigr)^{2\abs{\mathcal{L}^0}} ,
\end{align}
where $\abs{\mathcal{L}^0}$ denotes the dimension of the space $\mathcal{L}^0$, i.e.\ the number of zero eigenvalues.
Combining all contributions, we find
\begin{align}
\det\left(U^T\grad_x U\right)
={}& \prod_{\mathclap{k \in \mathcal{L}^+}}
\bigl[\sinc(\lambda_k/2)\bigr]^{2\abs{\mathcal{L}^0}} \cdot {} \notag \\*
&\prod_{\mathclap{k < l \in \mathcal{L}^+}}\bigl[\sinc\bigl((\lambda_l - \lambda_k)/2\bigr) \cdot {} \notag \\*
&\prod_{\mathclap{k < l \in \mathcal{L}^+}}\sinc\bigl((\lambda_l + \lambda_k)/2\bigr)\bigr]^{\mathrlap{2}} .
\end{align}
We get a more compact expression by letting the indices roam over the full set and compensating with a square root for the double product, which leads to
\begin{equation}
\det\left(U^T\grad_x U\right)
= \prod_{k<l}\sinc\biggl(\frac{\lambda_l + \lambda_k}{2}\biggr) .
\end{equation}


\section{Jacobian of the Givens Rotations}\label{sec:detjac_givens}

For parametrizaiton via Givens rotations in the following order
\begin{align}
U &= \sideset{}{^\downarrow}\prod_{k<l}G(k,l,x_{kl}) \notag \\
&= G(n-1,n,x_{n-1,n})\dotsb G(1,2,x_{12}),
\end{align}where $G(k,l,x_{kl})$ is the rotation matrix in the $k^{th},l^{th}$ plane with angle $x_{kl}$ and $\prod^{\downarrow}$ indicates a product in descending order (reading from left to right) i.e.\ $kl= (N-1)N, (N-2)N,\dotsc,23,13,12$. Note that the demonstration would also hold for matrices ordered in ascending order. We denote $G^<(i,j)\isDefinedAs\prod^{\downarrow}_{(kl) < (ij)}G(k,l,x_{kl})$ and $G^>(i,j)\isDefinedAs\prod^{\downarrow}_{(kl) > (ij)}G(k,l,x_{kl})$ so that $U = G^<(i,j)G(i,j,x_{ij})G^<(i,j)\; \forall i<j$. A simple matrix product gives $G^T(i,j,x_{ij})\deriv{G(i,j,x_{ij})}{x_{ij}}=\deltaMinus_{ji}$, so 
\begin{align}
    \left(U^T\deriv{U}{x_{ij}}\right)_{\mathrlap{pq}}
    &= \Big(G^{<T}(i,j)G^T(i,j,x_{ij}) \notag\\*
    &\hphantom{{}=\Bigl(}\deriv{G(i,j,x_{ij})}{x_{ij}}G^{<}(i,j) \Big)_{pq} \notag \\
    &= G^{<T}_{pj}(i,j)G^{<}_{iq}(i,j) \notag \\*
    &\hphantom{{}={}}{}- G^{<T}_{pi}(i,j)G^{<}_{jq}(i,j).
\end{align}
Since $i<j, U^T\deriv{U}{x_{ij}}$ is actually an upper block triangular matrix in which the lower part $pq$ is zero for $q > j$
\begin{align*}
\kbordermatrix{pq\diagdown ij & 12 & 13 & 23 & 14 & 24 & 34 & \ldots \\
12 & -1 & * & * & * & * & * & \ldots \\
13 &   & * & * & * & * & *  & \ldots \\
23 &   & \Box & * & * & * & * & \ldots \\
14 &   &  &  & * & * & * & \ldots \\
24 &   &  &  & \Box & * & * & \ldots \\
34 &   &  &  & \Box & \Box & * & \ldots \\
\vdots & & & & & & & \ddots
} .
\end{align*}
To triangularize the matrix further, note that each diagonal block $j$ (so $ij$ with $i < j$) starts with \(G^{<T}_{pq}(1,j) = \bigl(G^T(1,2) \dotsb G^T(j-2,j-1)\bigr)_{pq}\), which can be eliminated by multiplying each block from the left by $G^<$, we will introduce
\begin{align}
\G^< = \begin{pmatrix}
1 \\
& G_{2\times 2}^<(1,3) \\
&& G_{3\times3}^<(1,4) \\
&&& \ddots
\end{pmatrix}
\end{align}
or expressed in components
\begin{equation}
    \G^<_{pq,rs}(k,l)= \delta_{qs} G_{pr}^<(1,q).
\end{equation}
Since this is just a sequence of Givens rotations, we directly have \(\det(\G^<) = 1\), so the determinant is not changed.
The diagonal blocks now reduce to
\begin{align}
    \left(\G^<U^T \deriv{U}{x_{ij}}\right)_{\mathrlap{pj}} &=
    \left(\prod_{k<i}G(k,j)\right)_{\mathrlap{jp}}G^{<}_{ij}(i,j) \notag \\
    &- \left(\prod_{k<i}G(k,j)\right)_{\mathrlap{ip}}G^{<}_{jj}(i,j)
\end{align}
The first term on the r.h.s.\ vanishes, since
\begin{align}
G^<_{ij}(i,j) &= \sum_r\bigl(G(i-1,j)\dotsb G(1,j)\bigr)_{ir} \notag \\*
&\hphantom{{}={}\sum_r\qquad} G^<_{rj}(j-2,j-1) \notag \\
&= \bigl(G(i-1,j)\dotsb G(1,j)\bigr)_{ij} = 0 ,
\end{align}
where we used that $i < j$. 
For the product over $k < i$ in the second term we have
\begin{align}
\bigl(G(i-1,j) \dotsb G(1,j)\bigr)_{ip} = \delta_{ip}
\end{align}
and the diagonal of $G^{<}(i,j)$ yields
\begin{align}
G^{<}_{jj}(i,j)
&= \bigl(G(i-1,j)\dotsb G(1,j)\bigr)_{jj} \notag \\
&= \cos(x_{i-1,j})G^{<}(i-1,j)_{jj} + {} \notag \\*
&\hphantom{{}={}}
\sin(x_{i-1,j})G^{<}(i-1,j)_{j-1j} \notag \\
&=\prod_{k=1}^{i-1}\cos(x_{kj})
\end{align}
by induction and we used that $G^{<}(i-1,j)_{j-1j}  = 0$.
So for the determinant we find
\begin{multline}
    \det\left(U^T\grad_x U\right) = \prod_{i<j}(-1)\prod_{k=1}^{i-1}\cos(x_{kj}) \\
    = (-1)^{N(N-1)/2}
    \prod_{l\geq3}^N\prod_{k=1}^{l-2}\cos(x_{kl})^{l-k-1}.
\end{multline}

\section{Jacobian of the Householder reflections}\label{sec:detjac_householder}

In this Section we take $U=\prod_{l}^{\downarrow}H(l) = H(N) \dotsb H(2)$, with $H(l)$ Householder reflection w.r.t.\ to the plane orthogonal to $v_l$, defined in~\eqref{eq:householder_vect_def} (we take this ordering of the product for simplicity, but the proof works analogously for any ordering). Note that, by construction, $H(l)v_l=-v_l$.
\begin{align}
    H^T(l)\deriv{H(l)}{x_{ij}} &=-\frac{2\delta_{lj}}{\norm{v_l}^2}H(l) \Big(e_i v^T_l + v_l e_i^T \notag \\
    &\hphantom{{}={}}{} +x_{li}(H(l)-1)\Big) \notag \\
    &=-\frac{2\delta_{lj}}{\norm{v_l}^2}(e_iv_l^T-v_l e_i^T).
\end{align}
Similarly to the case of the Givens rotations, we define $H^<(l)\isDefinedAs\prod^{\downarrow}_{k<l} H(k)$ and $H^>(l)\isDefinedAs\prod^{\downarrow}_{k>l} H(k)$, with the convention that \(H^<(2) = H(1) = 1\)
\begin{align}
    \left(U^T\deriv{U}{x_{ij}}\right)_{\mathrlap{pq}}
    &=\left(H^{<T}(j)H^T(j)\deriv{H(j)}{x_{ij}}H^{<}(j)\right)_{pq} \notag \\
    &=-\frac{2}{\norm{v_j}^2}\Big(H^{<T}_{pi}(j)\big(v_j^TH^{<}(j)\big)_q \\*
    &\hphantom{{}=-\frac{2}{\norm{v_j}^2}\Big(}
    -\big(H^{<T}(j)v_j\big)_p H_{iq}^{<}(j)\Big). \notag 
\end{align}
$(v_j^T H^{<}(j))_q= v_{qj}=0$ for $q>j$ and $H^{<}_{iq}(j)=0$ for $q\geq j ( > i)$ so again $U^T\grad_x U$ is upper block triangular. For the blocks on the diagonal, $q=j$,
\begin{equation}
    \left(U^T\deriv{U}{x_{ij}}\right)_{\mathrlap{pj}} = -\frac{2}{\norm{v_j}^2}H^{<T}_{pi}(j).
\end{equation}
So the matrix actually has the shape
\begin{align*}
-2
\begin{pmatrix}
\frac{1}{\norm{v_2}^{2}} & {*} & {*} & \ldots \\
 {0}  &\frac{H^{<T}_{2\times2}(3)}{\norm{v_3}^{2}} & {*} & \ldots\\
 {0}  &{0}  &\frac{H^{<T}_{3\times3}(4)}{\norm{v_4}^{2}} & \ldots\\
 \vdots & \vdots & \vdots & \ddots
   \end{pmatrix}
\end{align*}
The determinant is thus simply the product of the determinants of each block
\begin{align}
    \det(U^T\grad_x U)&=\prod_{l=2}^N \det\bigl(-2\norm{v_l}^{-2}H^{<T}(l) \bigr) \notag \\
    &= 2^{N(N-1)/2}\prod_{l=2}^N\norm{v_l}^{2(1-l)} ,
\end{align}
where we used that $\det\big(H(l)\big)=-1\;\forall l$.

\section{Conservation of the Uniqueness of the Minimum by Composition with Periodic Function with a Unique Minimum}\label{sec:unic_min_periodic}

We consider $n\circ f$ with $n$ having only one minimum and $f$ periodic with one minimum per period, two real-valued continuous functions defined on a subset of $\mathds{R}$. $n\circ f$ will be periodic of the same period as $f$ and we first restrict the reasoning to one period. There exist $I^-$ and $I^+$ domains (possibly  empty) on which $n$ is respectively non-increasing and non-decreasing s.t.\ \(I^- \leq I^+\), i.e.\ \(\forall f_i\in I^-\) and \(\forall f_j\in I^+\) we have \(f_i\leq f_j\) (with \(I^-\cup I^+\), the domain of definition of $n$), 
and similarly, there exist $X^-$ and $X^+$ for which $f$ is respectively non-increasing and non-decreasing s.t.\ \(X^- \leq X^+\) (by shifting the period to start at the unique maximum of $f$). Then $n\circ f$ is non-increasing on $x \in X^{\mp}$ for $f(x)\in I^{\pm}$ and non-decreasing on $x\in X^{\pm}$ for $f(x)\in I^{\pm}$ (recapitulated in Table \ref{tab:behav_Ef}). 
\begin{table}
    \begin{center}
    \caption{Monotonicity of $n\circ f$ (non-decreasing $\nearrow$ or non-increasing $\searrow$) depending on the domain of $x$ and $f(x)$}
    \label{tab:behav_Ef}
    \begin{tabular}{c|cccc}
        $x\in$& $X^-$ & $X^-$ & $X^+$ & $X^+$ \\
        $f(x)\in$ & $I^+$ & $I^-$ & $I^-$ & $I^+$ \\
        \hline
       $n\circ f$ & $\searrow$ & $\nearrow$ & $\searrow$ & $\nearrow$
    \end{tabular}
    \end{center}
\end{table}

So, either $Imag(f)\subseteq I^+$ and $n\circ f$ have a single minimum at $x\in X^-\cap X^+$, or $Imag(f)\subseteq I^-$ and $n\circ f$ as a minimum at the boundaries, which is equal by periodicity, or $n\circ f$ has a minimum for $f(x)\in I^-\cap I^+$, i.e.\ the unique minimum of $n$. So $n$ has a unique minimum per period, which is a degenerate minimum, but other (i.e.\ local) minima.

\section{Example of an Artificial Critical Point for Two-Electrons and Two Orbitals}\label{sec:crit_pt_ex}

Let us consider a 2-electron system, with 2 orbitals and to have simple expressions, we take a specific Hamiltonian. For the 1-electron part, we take
\begin{equation}
    h=\frac{1}{2}\begin{pmatrix}
        3 & -5 \\
        -5 & -9
    \end{pmatrix}
\end{equation}
and for the 2-electron integrals $[11|11]=[22|22]=2$ and $[11|12]=[11|22]=[12|12]=[12|22]=1$. We consider the Müller functional, defined in~\eqref{eq:muller_func} and the Givens parametrisation (equivalent to the exponential one for 2 orbitals). The energy function w.r.t.\ the parameters is
\begin{equation}
    \begin{split}
    \Enx&(n_1,n_2,x_{12}) = \frac{1}{8}\Big(9(n_1^2+n_2^2)-21(n_1+n_2) \\
    &+6n_1n_2-6\sqrt{n_1}\sqrt{n_2} +24(n_1-n_2)\cos(2x_{12})\\
    &+\big((\sqrt{n_1}-\sqrt{n_2})^2-(n_1-n_2)^2)\big)\cos(4x_{12}) \\
    &+\big(8(n_1^2-n_2^2)-16(n_1-n_2)\big) \sin(2x_{12})\\
    &-(\sqrt{n_1}-\sqrt{n_2})^2\sin(4x_{12})\Big).
\end{split}
\end{equation}
For only 2 orbitals, the only possibility to get a critical point due to the parametrization is $n_1=n_2$, the gradient is then
\begin{align}
    &\grad_{n,x}\Enx\vert_{n_1=n_2}=\notag\\
    &\begin{pmatrix}
        3\cos(2x_{12}) + (n_1-1)(2\sin(2x_{12})+3) \\
        -3\cos(2x_{12}) - (n_1-1)(2\sin(2x_{12})-3) \\
        0
    \end{pmatrix}.
\end{align}
So for $n_1=n_2=1$ and $x_{12}=\pi/4[\pi/2]$, $\grad_{n,x} E=0$, which corresponds to the 1-RDM
\begin{equation}
    \gamma^* = \begin{pmatrix}
        1 & 0 \\
        0 & 1
    \end{pmatrix}. 
\end{equation}
However $\grad_{\gamma}E\vert_{\gamma=\gamma^*}=(3,0,-3)^T\neq 0$ so this gives an example of critical point for $\Enx$, which is not a critical point of $E$. Now, we can verify the shape of the null space
\begin{align}
    &\grad_{n,x}\gamma=\notag\\
    &\begin{pmatrix}
        \cos(x_{12})^2 & \sin(x_{12})^2 & (n_2-n_1)\sin(2x_{12})\\
        \frac{1}{2}\sin(2x_{12}) & -\frac{1}{2}\sin(2x_{12}) & (n_1-n_2)\cos(2x_{12})\\
        \sin(x_{12})^2 & \cos(x_{12})^2 & (n_1-n_2)\sin(2x_{12})
    \end{pmatrix}
\end{align}
which at $n_1=n_2(=1)$ and $x_{12}=\pi/4[\pi/2]$ reduced to 
\begin{equation}
    \grad_{n,x}\gamma=\frac{1}{2}\begin{pmatrix}
        1 & 1 & 0\\
        1 & -1 & 0 \\
        1 & 1 & 0
    \end{pmatrix}.
\end{equation}
We clearly see that an artificial critical point will appear for $\grad_{\gamma}E$ of the form $(a,0,-a)^T$ (with $a\in\mathds{R}$), the case at $\gamma=\gamma^*$.


\bibliography{bibliography.bib}

\end{document}